\documentclass[12pt,preprint]{aastex}
\shorttitle{Parsec-scale radio imaging of NGC~6764}
\shortauthors{Kharb et al.}
\begin{document}
\title{Parsec-scale Imaging of the Radio-bubble Seyfert galaxy NGC~6764}
\author{P. Kharb\altaffilmark{}}
\affil{Department of Physics, Rochester Institute of Technology, Rochester, NY 14623, USA}
\email{kharb@cis.rit.edu}
\author{Ananda Hota\altaffilmark{}}
\affil{Institute of Astronomy and Astrophysics, Academia Sinica, Taiwan}
\author{J. H. Croston\altaffilmark{}}
\affil{School of Physics and Astronomy, University of Southampton, UK}
\author{M. J. Hardcastle\altaffilmark{}}
\affil{Centre for Astrophysics Research, University of Hertfordshire, UK}
\author{C. P. O'Dea\altaffilmark{}}
\affil{Department of Physics, Rochester Institute of Technology, Rochester, NY 14623}
\author{R. P. Kraft\altaffilmark{}}
\affil{Harvard Smithsonian Center for Astrophysics, 60 Garden St, Cambridge, MA, USA}
\author{D. J. Axon\altaffilmark{}}
\affil{Department of Physics, Rochester Institute of Technology, Rochester, NY 14623}
\affil{Department of Physics and Astronomy, University of Sussex, UK}
\and
\author{A. Robinson\altaffilmark{}}
\affil{Department of Physics, Rochester Institute of Technology, Rochester, NY 14623}

\begin{abstract}
We have observed the composite active galactic nucleus (AGN)-starburst galaxy NGC~6764 with the Very Long Baseline Array at 1.6 and 4.9~GHz. These observations have detected a ``core-jet'' structure and a {\it possible} weak counterjet component at 1.6~GHz. The upper limits to the core and jet (1.6$-$4.9~GHz) spectral index are 0.6 and 0.3, respectively. Taken together with the high brightness temperature of $\sim10^{7}$~K for the core region, the radio emission appears to be coming from a  synchrotron jet. At a position angle of $\sim25\degr$, the parsec-scale jet seems to be pointing closely toward the western edge of the southern kpc-scale bubble in NGC~6764. A real connection between the parsec- and sub-kpc-scale emission would not only suggest the presence of a curved jet, but also a close link between the AGN jet and the radio bubbles in NGC~6764. We demonstrate that a precessing jet model can explain the radio morphology from parsec- to sub-kpc scales, and the model best-fit parameters of jet speed and orientation are fully consistent with the observed jet-to-counterjet surface brightness ratio. The jet however appears to be disrupted on scales of hundreds of parsecs, possibly due to interaction with, and entrainment of the interstellar medium gas, which subsequently leads to the formation of bubbles. 
The jet energetics in NGC~6764 suggest that it would take 12$-$21~Myr to inflate the (southern) bubble. This timescale corresponds roughly to the starburst episode that took place in NGC~6764 about 15$-$50~Myr ago, and could be indicative of a close connection between jet formation and the starburst activity in this galaxy.
\end{abstract}

\section{Introduction}
Radio emission in Seyfert galaxies comes typically from weak outflows emanating from the active galactic nucleus (AGN), which appear to be emitted in random directions with respect to the host 
galaxy plane \citep{Kinney00,Gallimore06}. While the outflows in some Seyfert galaxies can extend 
to sub-kpc or even kpc-scales \citep{Ulvestad84,Baum93,Colbert96}, they are rarely in the form of the type of collimated (100-)kpc-scale jets that are observed in radio-loud AGNs. Sensitive radio observations have, however, revealed the presence of bubble-like structures in some Seyfert galaxies \citep{Elmouttie95,Irwin03, HotaSaikia06,Kharb06}. 
While starburst-driven superwinds have been invoked to explain the radio bubbles in Seyferts   
 \citep[e.g.,][]{Baum93}, an AGN-related origin has been favored by other studies
\citep{HotaSaikia06,Wilson81,Wehrle88,Taylor92}.
Copious amounts of X-ray emission, usually explained as thermal radiation from hot gas, are found to be closely associated with the largely non-thermal radio emission \citep{Cecil95,Wilson01,Croston08}. The complex radio and X-ray morphologies observed in Seyfert galaxies suggest that the emission may come from a number of different, yet closely related, mechanisms.

On parsec scales, Very Long Baseline Interferometry (VLBI) observations typically reveal 
weak radio ``cores'' and ``jet-like'' extensions in Seyfert galaxies \citep{Middelberg04}. 
While Doppler boosting of emission has been invoked in a few Seyfert jets \citep[e.g., Mrk~231;][]{Reynolds09}, Seyfert jets are normally sub-relativistic \citep{Ulvestad98}. A link between star formation and Seyfert nuclear activity has often been suggested \citep[e.g.,][]{Weedman83,Davies07}. On account of the faintness of the AGN emission, contributions from nuclear starbursts can become significant, thereby confusing the exact origin of the radio emission in Seyfert galaxies.

NGC~6764 is a nearby ($z$=0.00806\footnote{At the distance of NGC~6764, $1\arcsec$ 
corresponds to 154 pc, for $H_0$ = 71 km~s$^{-1}$~Mpc$^{-1}$, $\Omega_{mat}$ = 0.27, and $\Omega_{vac}$ = 0.73.}, luminosity distance = 32.2 Mpc) barred 
spiral galaxy (type SB(s)bc), alternatively classified as a LINER, a Seyfert 2, and a Wolf-Rayet galaxy (one with signatures of recent massive star formation). Kiloparsec-scale radio observations with the Very Large Array (VLA) and the Giant Meterwave Radio Telescope (GMRT) have revealed $\sim1$ kpc (total extent $\sim2.6$ kpc) north-south (N-S) orientated bipolar bubbles of non-thermal radio emission along the minor axis of the spiral host galaxy \citep{HotaSaikia06}. In addition, the higher resolution 5~GHz and 8~GHz VLA A-array configuration images of \citet{HotaSaikia06} show a region of bright radio emission extending roughly in the east-west (E-W) direction.

The {\it ROSAT} X-ray Observatory detected a highly variable X-ray source in NGC~6764, which
pointed to the presence of a compact, actively accreting AGN \citep{Schinnerer00}. Recent {\it Chandra} X-ray observations of NGC~6764 have discovered extended X-ray emission, well coincident with the N-S orientated radio bubbles \citep{Croston08}. Bright X-ray emission is also observed in the E-W 
region coincident with the high surface brightness radio emission. X-ray spectral fitting of the N-S bubbles has shown that the X-ray data are best fitted by a hot thermal gas model with $k$T = 0.75 keV, while the E-W extension is best fitted by hot thermal gas plus absorption with $k$T = 0.93 keV and 
$N_H = 2 \times 10^{21}$ cm$^{-2}$. The total energy stored in the hot gas is high ($\sim10^{56}$ erg), and cannot be ascribed to a nuclear starburst alone; an AGN-related origin appears much more likely \citep[see][]{Croston08}. 

The X-ray hardness ratio maps (using the 1.0$-$5.0 keV and 0.4$-$1.0 keV filtered images) indicate that the western edge of the kpc-scale bubbles has a relatively hard X-ray spectrum. Radio spectral index maps (using the 1.4$-$4.9 GHz VLA and the 1.4$-$8.4 GHz VLA and GMRT images) revealed the south-western edge to have a flatter radio spectrum \citep[see][]{HotaSaikia06}. 
The radio and X-ray spectral flattening could be signifying enhanced localized particle reacceleration
along the south-western edge of the bubbles. The radio$-$X-ray results strongly support a scenario wherein an AGN outflow is shock heating and entraining the warm interstellar medium (ISM)
gas in the south-western region of the bubble. However, the jet direction
could not be unambiguously determined from the kpc-scale radio observations \citep{HotaSaikia06}, leading to some confusion about how the radio/X-ray bubbles were powered 
by the AGN. In order to resolve this issue, we observed NGC~6764 with the Very Long Baseline 
Array (VLBA) at 1.6 and 4.9 GHz, and we report the results of these observations in this paper.

\section{VLBA Observations}
We observed NGC~6764 with the ten elements of the VLBA \citep{Napier94}
on 2008 November 13, at 4.9 and 1.6 GHz (Project ID BK154), in a phase-referencing experiment \citep{BeasleyConway95}. The calibrator 1758+388 was used as the 
fringe-finder, while the compact source 1917+495 with a switching angle of 2$\degr$ 
and an X, Y, positional error of 0.38, 0.70 $mas$, respectively, was chosen as the phase reference calibrator\footnote{See http://www.vlba.nrao.edu/astro/calib/}. A switching cycle 
of 5 minutes (calibrator 2 minutes and target 3 minutes) was adopted at both the frequencies. The 
calibrator 1924+507 ($\sim2.8\degr$ away and with a high positional accuracy) was
used as the phase check source. A phase check source could be used to get a
measure of the decorrelation expected on the target, in the case of a non-detection. 

\begin{figure}
\centerline{
\includegraphics[width=9cm]{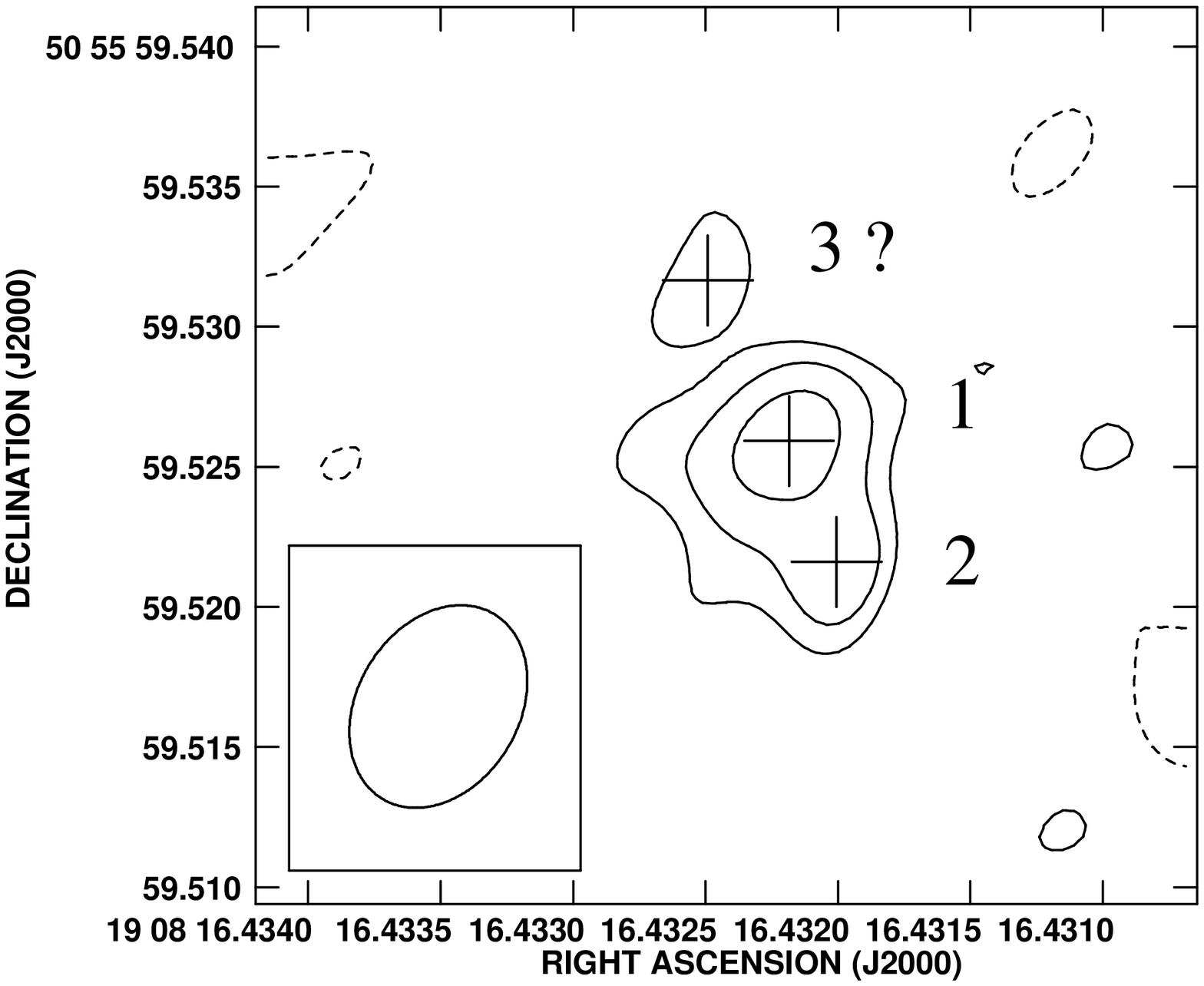}
\includegraphics[width=9cm]{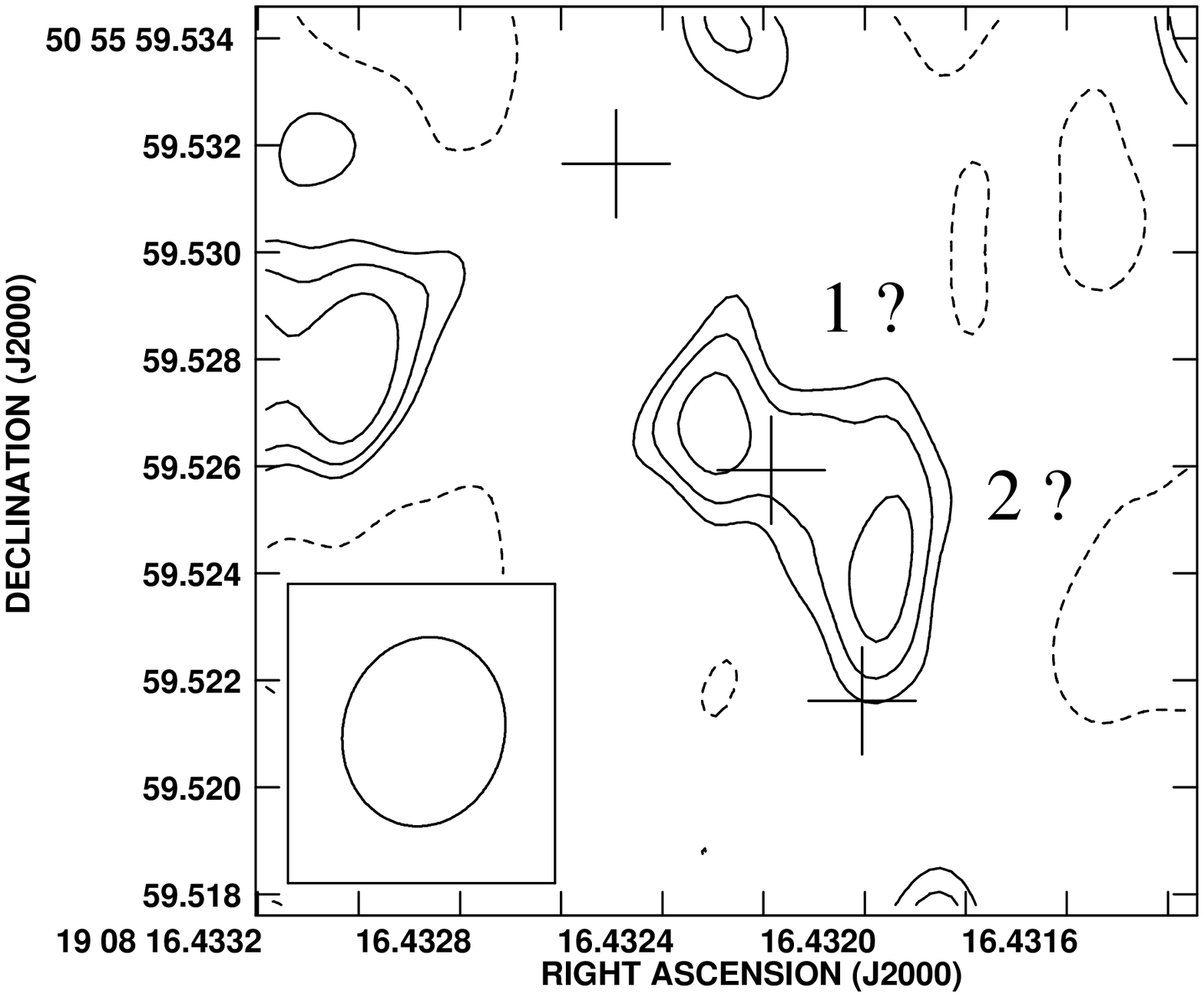}}
\caption{1.6 GHz (left) and 4.9 GHz (right) surface brightness contour maps of the parsec-scale structure in NGC~6764. The contours are 
(left) $\pm$0.18, 0.26, 0.37 mJy~beam$^{-1}$, and
(right) $\pm$0.09, 0.13, 0.18 mJy~beam$^{-1}$.
We identify the components 1, 2, and 3, to be the core, jet, and a {\it possible} counterjet. The peak positions of the three components observed at 1.6~GHz have been marked with crosses in both the panels.}
\label{fig:vlbi}
\end{figure}

The observations\footnote{http://www.vlba.nrao.edu/astro/VOBS/astronomy/nov08/bk154/} were made using the standard setup with 128~Mbps recording and a 2-bit sampling rate, with two 8~MHz intermediate frequency channels (IFs) in dual polarization. This yielded a total bandwidth of  ($8\times2\times2=$) 32~MHz.
We followed the standard VLBA data reduction procedures using the NRAO 
Astronomical Image Processing System \citep[AIPS,][]{Greisen03} as described in 
Appendix C of the AIPS Cookbook. Delays were corrected using fringe fitting on the 
fringe finder (with AIPS task FRING). The phase calibrator 1917+495 was iteratively 
imaged and self-calibrated using the AIPS tasks IMAGR and CALIB. The images were then 
used as models to determine the amplitude and phase gains for the antennas. These 
gains were applied to the target and phase check source, and images were made using 
IMAGR. Due to the faintness of the source, self-calibration was not possible (there were too many failed solutions).
The total on-source time was $\sim$120~minutes at 1.6 GHz, and $\sim$150~minutes at
4.9 GHz, which resulted in a final $rms$ noise of $\sim90~\mu$Jy~beam$^{-1}$
at 1.6 GHz, and $\sim80~\mu$Jy~beam$^{-1}$ at 4.9 GHz.
The images displayed in Figure~\ref{fig:vlbi} were created with uniform weighting 
(with ROBUST 0 in IMAGR for 1.6~GHz, and ROBUST 5, UVTAPER 0 $-$ 70 M$\lambda$ for 4.9 GHz). The insets display the elliptical Gaussian beams with
dimensions = $7.7~mas\times5.7~mas$, at a position angle (PA) of $-31\degr$ for 1.6~GHz,
and $3.6~mas\times3.0~mas$ at a PA of $-13\degr$ for 4.9~GHz.

\section{Results}
We have detected two distinct radio components at 1.6 GHz (marked 1 and 2 in Figure~\ref{fig:vlbi}). 
There is a suggestion of a weak third component to the north of component 1. 
New sensitive radio observations are required to confirm this feature. 
The source position is $\sim0.58\arcsec$ away 
from the listed optical host galaxy position of R.A. 19h 08m 16.370s, decl. 50d 55m 59.58s  \citep{Clements81}.
We identify components 1, 2 and 3 to be the core, the jet, and a {\it possible} counterjet,
respectively. The ``core-jet'' structure extends to $\sim$0.8 parsec, at a P.A. of $\sim25\degr$. The peak surface brightness of the core, and the total radio 
flux density of the ``core-jet'' structure are 0.46 mJy~beam$^{-1}$ and 
0.80 mJy, respectively. The peak intensities and positions of the radio components, estimated
using AIPS tasks JMFIT and IMDIST, are listed in Table~\ref{tabparam}. 

\begin{deluxetable}{ccclccc}
\tabletypesize{\small}
\tablecaption{Source Parameters at 1.6~GHz}
\tablewidth{0pt}
\tablehead{
\colhead{Comp.} & \colhead{Peak} & \colhead{Total} & \colhead{Position}& \colhead{Angular} & \colhead{Linear}\\
\colhead{} & \colhead{Intensity} & \colhead{Intensity} &\colhead{R.A., Decl.} & \colhead{Separation} & \colhead{Separation}}
\startdata
1 & 0.46$\pm$0.09 & 0.50$\pm$0.18 & 19h 08m 16.4322s, 50$\degr$ 55$\arcmin$ 59.526$\arcsec$&.... &.... & \\ 
2 & 0.35$\pm$0.09 & 0.40$\pm$0.16 & 19h 08m 16.4320s, 50$\degr$ 55$\arcmin$ 59.522$\arcsec$&4.3$\arcsec$ &0.7 & \\
3 & 0.23$\pm$0.09 & 0.30$\pm$0.20 & 19h 08m 16.4325s, 50$\degr$ 55$\arcmin$ 59.532$\arcsec$&6.1$\arcsec$ &0.9 & \\
\enddata
\tablecomments{Column~1: component number. Columns~2 and 3: peak and integrated intensity in mJy~beam$^{-1}$.
Column~4: position of components in sky. Columns~5 and 6: the angular and linear projected separation between components
2 and 3 with respect to component 1, in arcseconds and parsecs, respectively.}
\label{tabparam}
\end{deluxetable}

In order to test the credibility of component~3, we estimated a surface density of spurious noise peaks in the image by dividing the number of noise peaks having a signal-to-noise ratio (S/N) similar to component~3 ($i.e.,$ S/N~$>$2.6), by the entire image. We found around 230 noise peaks in an image of size $0.4\arcsec\times0.4\arcsec$, resulting in a noise peak surface density of $\sim1370$ peaks arcsec$^{-2}$. Considering then a region of size ($15~mas\times15~mas$), centered around the peak
source emission, as the region where a noise peak could be mistaken for source
emission (the core-jet distance being $\sim5~mas$), we estimated that 0.3 noise
peaks could be expected in this region. Therefore, there is a 30\% chance that component 3 is a noise peak. 

The 1.4~GHz peak flux density of the VLA A-array core (size $\sim1.5\arcsec$) is 
$\sim14$~mJy \citep{HotaSaikia06}. This
implies that only about 6\% of the VLA flux density is detected by the VLBA.
This is similar to what is observed in the Seyfert galaxy NGC~4151 \citep[$\sim$8\%;][]{Pedlar93,Ulvestad98}. 
We believe, however, that had self-calibration 
worked in NGC~6764, its fraction of VLBA to VLA flux density would have been higher.
Nevertheless, it appears that there is either a lot of diffuse emission on scales of tens or hundreds of parsecs which 
are not visible to the VLBA, or the VLBA is not sensitive to the diffuse radio emission on parsec scales (this appears 
to be less likely in the case of NGC~6764 considering the large fraction of missing flux), or a combination of both 
(see Orienti \& Prieto 2010 for a discussion on the missing diffuse emission in VLBI observations).  

There appears to be a tentative detection of components 1 and 2 at 4.9 GHz. However, this 
detection is also in need of new confirmatory observations, as the radio peak in components 
1 and 2 is only around 3 times the $rms$ noise ($\sim80~\mu$Jy~beam$^{-1}$), while the 
total flux density is only $\sim$0.3~mJy. As the 4.9 GHz emission is weak,
we have not attempted to create a spectral index map. Using the $3\times${\it rms} value
from the 4.9 GHz map, we estimate upper limits to the spectral indices (defined such that 
$S_{\nu}\propto\nu^{-\alpha}$, where $S_{\nu}$ is the flux density at frequency $\nu$). These are 
$>0.6$ and $>0.3$ for components 1 and 2, respectively. 

The steep radio spectrum of the radio core is consistent with many Seyfert core observations in 
the literature \citep[e.g.,][]{Sadler95,Roy00,Orienti10}, and suggests optically-thin synchrotron emission. The steep radio spectrum of the radio core rules out thermal free-free radiation as the dominant emission mechanism. The brightness temperature of 
the radio emission ($T_{B}$) can be estimated using the relation, 
$T_{B}=1.8\times10^{9}~(1+z)~(\frac{S_{\nu}}{1~mJy})~(\frac{\nu}{1~GHz})^{-2}~(\frac{\theta_{1}\theta_{2}}{2~mas^{2}})^{-1}~K$, 
where $z$ is the redshift, $\theta_{1}, \theta_{2}$ are, respectively, the major and minor axes of the beam, and
the factor 1/2 arises for the case of ``unresolved'' components \citep{Ulvestad05}.
For a core region of size (= beam-size) = $7.7~mas\times5.7~mas$, the brightness temperature is of the order of $1.5\times10^7$~K. This is consistent with the brightness temperatures observed in other Seyfert nuclei \citep{Kukula99,Middelberg04, Orienti10}. The high brightness temperature provides further support to a non-thermal origin of the radio emission. We note that starburst galaxies typically 
exhibit brightness temperatures of $<10^{5}$~K \citep{Condon91}. 

\section{Discussion}
At a P.A. of $\sim25\degr$, the parsec-scale jet is pointing closely to the western edge of 
the southern kpc-scale radio bubble. This region (marked ``A'' in the top panel of Figure~2) has a flatter radio and a harder X-ray spectrum \citep{HotaSaikia06, Croston08}. The VLBA jet appears to be in the same direction 
as the weak jet-like extension observed in the high-resolution VLA (A-array, 8 GHz) observations, which is marked ``Jet?'' in Figure~11 of \citet{HotaSaikia06} and in the middle panel of Figure~2. A connection between the radio emission on parsec- and sub-kpc scales in NGC~6764 would require the jet to be curved. While this connection needs to be examined in greater detail with
more sensitive sub-kpc resolution (e.g., 5~GHz EVLA A-array) radio observations, bent or S-shaped radio jets have often been observed in Seyfert galaxies \citep[e.g.,][]{Kukula95,Nagar99,Thean00}. Three popular scenarios for jet bending have been expounded: (1) precession of the jet ``nozzle'' in the galactic nucleus \citep{Ekers78}, (2) ram-pressure of the rotating ISM \citep{Wilson82}, and (3) static pressure gradients in the hot gaseous galactic halo \citep{SmithNorman81}. However, a combination of these and possibly other effects (like starburst superwinds bending the slow radio jets, as suggested for NGC~6764 by \citet{HotaSaikia06}) could all be in play here, to varying extents. It is also feasible that the misaligned VLBI jet is a new jet which has little to do with the N-S bubbles, which are in turn, just relic emission from a previous activity episode. We briefly explore these ideas ahead.

\subsection{Precessing Jet and Bubbles}
Precessing radio jets have been directly observed in Galactic X-ray binaries \citep[e.g.,][]{Mioduszewski01} and suggested to be present in Seyfert galaxies with curved jets like NGC~3516 \citep{Veilleux93}. The detection of two pairs of orthogonally-placed, self-similar edge-brightened radio bubbles in the Seyfert galaxy Mrk~6, prompted \citet{Kharb06} to invoke two episodes of precessing radio jets to explain the radio structures. Jet precession could arise due to the presence of binary black holes \citep{Caproni04}, the relativistic Lense-Thirring effect \citep{Bardeen75,Caproni06}, or accretion disk warping due to non-axisymmetric radiation pressure forces \citep{Pringle96}. The Seyfert galaxy Circinus has radio bubbles with bright, highly polarized edges \citep{Elmouttie95}, similar to the ``edge-brightened'' Mrk~6 and NGC~6764, and a precessing accretion disk, as evidenced by the VLBI water maser observations of \citet{Greenhill03}. A precessing disk and S-shaped radio jets with tightly correlated X-ray emission have also been observed in the Seyfert galaxy NGC~4258 \citep{Sanders82, Cecil00,Wilson01}. 
Therefore, the presence of a precessing jet in the radio-bubble galaxy, NGC~6764, is a viable possibility.

\citet{Hjellming81} presented a three-dimensional kinematic model to explain the proper motions of the precessing jet in the X-ray binary SS433. This model has since been successfully applied to the radio morphologies of radio-powerful AGNs \citep[e.g.,][]{Gower82}. Following the relations in \citet{Hjellming81}, we have attempted to fit a precessing jet model to the radio structure in NGC~6764. We find that this model is able to fit the radio structure from parsec to sub-kpc scales (see Figure~2). The best-fit model parameters are: jet speed = 0.028$c$, jet P.A. = 76$\degr$, 
inclination = 18$\degr$, precession cone half-opening angle = 3$\degr$, and angular velocity = 1.4$\times10^{-5}$ rad~yr$^{-1}$. Apart from the radio morphology, this model is consistent with a couple of different observational findings. In the top panel of Figure~2, ``A'' marks the region with the flatter radio and harder X-ray spectrum, while ``B'' marks the location of a prominent, curved $H{\alpha}$ filament \citep{Zurita00, HotaSaikia06,Leon07}, seen here as coincident with the precessing radio counterjet. 
\citet{HotaSaikia06} were the first to present and discuss the radio$-H{\alpha}$ image overlays.
The correspondence of the jet with these regions can explain both the spectral flattening in the south (particle re-acceleration) and emission lines in the north (shock excitation). Clearly, this study would greatly benefit from (shock-sensitive) [NII]/H$\alpha$ emission line observations of NGC~6764 \citep[e.g.,][]{Sharp10}.  

While keeping in mind the uncertainties involved in the surface brightness values of the faint detected features, we estimated the observed jet-to-counterjet surface brightness ratio ($R_{j}$) at three
positions along the jet and counterjet. We note that if component~3 is a noise peak, then the $R_{j}$ estimates above must be regarded as lower limits. Taking the peak surface brightness values of components 2 (jet)  and 3 (counterjet), we derive an $R_{j}\sim$1.5. However, the peak positions of components 2 and 3 are not equidistant from the peak of component 1 (core). Therefore, taking the peak position of component 2 as a reference, we obtained the surface brightness at a distance of $\sim4.5~mas$ from the core (this being the distance between components 1 and 2) in the counterjet direction. This resulted in $R_{j}\sim$1.9. Similarly, taking the peak position of component 3 as a reference, we obtained the surface brightness value at a distance of $\sim6~mas$ from the core (the distance between components 1 and 3) in the jet direction. This yielded an $R_{j}\sim$1.2.
This latter value is perhaps the more robust among the three estimates, as it relies on the ``peak'' surface brightness of the weak component~3, rather than on the noise around it (as was used to get $R_{j}\sim$1.9). This $R_{j}$ value of $\sim1.2$ matches closely with the expected jet-to-counterjet ratio 
($R_{j}=(\frac{1+\beta~cos~\theta}{1-\beta~cos~\theta})^p$) of $\sim1.15$, for a Doppler-boosted radio jet with speed 0.028$c$ and an inclination of 18$\degr$ (both precessing jet model best-fit parameters) for a jet structural parameter of $p\approx2+\alpha=2.6$ \citep[e.g.,][]{UrryPadovani95}. 
Therefore, the current parsec-scale data on NGC~6764 are {\it not inconsistent} with a precessing jet model. It is important to note that the half-opening angle of the precession cone
is only 3$\degr$, which implies that the jet is very gently curved. Low-pitch precessing jets have 
been proposed for radio-loud AGNs as well \citep{Conway93,Kharb10}.

However, there are potential difficulties in proposing a precessing jet model similar to that proposed in Mrk~6 by \citet{Kharb06}, for NGC~6764. In Kharb et al., we attempted to explain the entire edge-brightened bubble with the highly polarized edges, as the projection of a collimated precessing jet. This picture has definite shortcomings in the case of NGC~6764. As noted previously,
while part ``B'' of the (counter-) jet in Figure~2 may be coincident with a curved H$\alpha$ filament, indicative of the ionized hydrogen gas; there is no clear evidence of a jet-like extension in the region marked ``A'' in the top panel of Figure~2, beyond what is observed in the middle panel of Figure~2. The major clue to the presence of a jet there, are the flatter radio and harder X-ray spectra. Moreover, the H$\alpha$ image in \citet{Zurita00} shows other bright curved filaments, notably toward the south-eastern edge of the bubble, where a precessing jet model (with the same best-fit parameters as above) cannot be fitted. Lastly, a shock, rather than a collimated jet, could also produce the flatter radio and harder X-ray spectra.

Recent X-ray observations of central galaxies in clusters have indicated multiple X-ray cavities in some of them (e.g., Perseus~A, NGC~5813, A2626). The connection of these, sometimes non-collinear cavities, with the bent radio jets has suggested the picture of a precessing jet, or a jet that changes direction, inflating bubbles at multiple epochs \citep[see][]{Dunn06,Wong08,Falceta10,Randall10}. It is also possible, however, that the apparent differences in the direction are due to the dynamics of the intracluster medium. \citet{Sternberg08} have suggested that slow jets with larger opening angles could give rise to fat bubbles. 
Therefore, although the picture of precessing jets inflating bubbles and creating X-ray cavities in powerful cluster radio galaxies is yet to be fully established, it is tantalizing to suggest that a similar process could be taking place in Seyferts which have radio bubbles. This then 
suggests an intimate connection between radio bubbles and radio jets, and leads us naturally to the 
next topic of jet-ISM interaction and the inflation of bubbles.

\subsection{Jet-ISM Interaction and Bubbles}
As the entire radio emission in NGC~6764 cannot be fitted with a precessing or curved radio jet, disruption or de-collimation due to a strong interaction with the ISM or winds, from either the starburst or AGN, is indicated. As mentioned earlier, there is some evidence for direct interaction between the jet and the X-ray emitting plasma close to the nucleus.
Ram pressure bending due to a rotating ISM could also give rise to the jet curvature itself. Following an approach similar to \citet{Wilson82}, we tried to deduce if the radio jet in NGC~6764 bends in the correct sense, if it was indeed a result of jet$-$rotating-ISM interaction. Assuming that the spiral arms ``trail'' in spiral galaxies \citep[e.g.,][]{deVaucouleurs58}, we see that NGC~6764 must rotate counter-clockwise. Then the VLBA jet, at least, seems to ``lead'' the rotation of the galaxy and therefore, could be interacting with the rotating gas (see the bottom panel of Figure~2), although a possible problem for this model is that the jet on scales larger than that probed by the VLBA (as in the middle panel of Figure~2) would be moving in the opposite direction and against the rotating medium. Furthermore, the kpc-scale bubbles should be carried downstream in the direction of the rotating medium, which is not observed. Therefore, the jet$-$rotating-ISM argument becomes weak in NGC~6764.

An alternative model that can explain the absence of a collimated jet on sub-kpc scales is one in which the initially SW-NE jet strongly interacts with cold, dense material in the host galaxy \citep[e.g.,][]{Leon07,Middelberg07}, and is disrupted, giving rise to pressure-driven bubbles (perhaps containing hot thermal material) which expand to the north and south. The physics behind this phenomenon is that entrainment of cold material simultaneously reduces the bulk speed of the jet and increases (via dissipation) the internal energy density \citep[e.g.,][]{Begelman82}. Once the bulk speed of the jet is of the order of the transverse expansion speed driven by the internal energy density, one no longer has a jet, but a bubble. The bubbles would sweep up the gas at the leading edge and result in edge-brightening (due to increased thermal free-free emission), as is observed in NGC~6764.

One observational clue that might in principle go against the idea of bubbles sweeping up the gas, is that the X-ray emission in NGC~6764 seems to be fully coincident with the radio emission \citep[as is also observed in other Seyfert galaxies; see][]{Cecil95,Veilleux97,Wilson00}. An expanding superbubble should be accompanied by extended X-ray emission from hot gas, but not from the cooler shell, which should just exhibit H$\alpha$ emission. For the swept-up gas ahead of the shock-front that is ionized, the X-ray emission should be ahead of the radio emission. 

\subsection{Energetics}
Under the assumption of ``equipartition'' of energy between relativistic particles and the magnetic field 
\citep{Burbidge59,Miley80}, we estimated the magnetic field strength and other parameters 
for a cylindrical jet geometry. The total radio luminosity was estimated assuming that the radio spectrum extends 
from 10~MHz to 100~GHz with a spectral index of $\alpha$=0.6. Further, it was assumed that the relativistic protons 
and electrons have equal energies, and the radio emitting plasma has a volume filling factor of unity. From the 1.6~GHz image, we 
estimate the size of the radio jet to be $11.8~mas\times8.1~mas$. Following the relations in \citet{OdeaOwen87}, 
we obtain a total radio luminosity of $\sim$2.4$\times10^{37}$~erg~s$^{-1}$, minimum magnetic field ($B_{min}$) of 
$\sim$1.26 mG, minimum energy of $\sim$2.1$\times10^{49}$ erg, and a minimum pressure ($P_{min}$) of 
$\sim$1.5$\times10^{-7}$ dynes~cm$^{-2}$. 

The electron lifetime due to synchrotron radiative losses can be estimated using the relation,
$t_{syn}\simeq33.4~B_{10}^{-3/2}~\nu_{c}^{-1/2}$ Myr, where
$B_{10}$ is the magnetic field (``minimum'' $B$ field here) in units of 10 $\mu$G, and $\nu_{c}$ is the critical frequency in GHz \citep[e.g.,][]{Pacholczyk70}. 
This turns out to be $\sim$1.8$\times10^4$ yrs for a critical frequency of 1.6 GHz.
Assuming that the radio jet is being confined by, and is in pressure balance with the surrounding ISM gas ($i.e., P_{min}=nk_{B}T$), we estimate that the gas at a temperature of 0.25~keV would need to have a density of $\sim$400 cm$^{-3}$ in order to do so. 

\subsection{The AGN Jet and Starburst connection}
NGC~6764 is well understood to be a composite AGN-starburst system. From the {\it Chandra} X-ray
observations of NGC~6764, \citet{Croston08} have estimated the hot X-ray gas pressure of (4$-$7)$\times10^{-12}$~dynes~cm$^{-2}$ in the southern bubble. If we assume that the radio jet does $4PV$ work  in order to inflate this bubble, then for a bubble volume of $\sim1.46\times1.16\times1.16$~kpc$^{3}=1.93$~kpc$^{3}$, this turns out to be (9$-$16)$\times10^{53}$~ergs, assuming that the gas inside and outside the bubble is in a pressure balance. Further assuming that the efficiency ($\epsilon$) with which the total jet energy is tapped to produce radio luminosity is 1\% \citep[e.g.,][]{ODea85}, the jet with a radio luminosity of 2.4$\times10^{37}$~erg~s$^{-1}$ and a kinetic luminosity ($L_{rad}$/$\epsilon$) of $\sim2.4\times10^{39}$~erg~s$^{-1}$ would take 12$-$21 Myr to create the bubble.
Interestingly, this timescale matches closely with a starburst activity episode in NGC~6764. Spectral synthesis models of \citet{Schinnerer00} have suggested two major starburst episodes in the nuclear regions ($\sim3\arcsec$ = 460~pc) that occurred 3$-$5 Myr and 15$-$50 Myr ago. If the timescales are indeed correlated, it could suggest that an accretion event about 10$-$20 Myr ago might have initiated a starburst, and eventually powered the AGN jet \citep[e.g.,][]{Davies07,Tremblay10}. 
This then would be a strong indicator of a close connection between AGN activity and violent nuclear star formation in a Seyfert galaxy.

The presence of non-thermal filaments in some Seyfert galaxies has been suggested to be the aftermath of burst superbubbles \citep[e.g.,][]{Veilleux93, Veilleux97}. However, the fact that very often these filaments seem to pass through the AGN core \citep[e.g., NGC~3079;][]{Cecil01}, suggests strongly that bubbles, if present, are linked to the AGN, rather than to starburst or galactic superwinds alone. The results presented in this paper clearly highlight the close connection between an AGN outflow and the kpc-scale radio bubble in yet another Seyfert galaxy, NGC~6764.

Finally, it is interesting to note that, using a sample of 60 warm infrared galaxies, \citet{Kewley00} discovered a bimodal distribution in the compact 2.3~GHz radio luminosity: majority of the galaxies above and below a threshold 2.3~GHz luminosity of $L_{2.3}\sim1.7\times10^{21}$~W~Hz$^{-1}$ were AGNs and starbursts, respectively. We find that the 2.3~GHz luminosity of NGC~6764, which is $L_{2.3}\sim5.3\times10^{19}$~W~Hz$^{-1}$ (assuming a 1.6$-$2.3~GHz spectral index of 0.6), places it in the starburst category. However, the ``core-jet'' radio morphology, and the high brightness temperature, strongly supports the presence of an AGN in NGC~6764.  

\section{Summary and Conclusions}
\begin{enumerate}
\item[1.] We have observed the composite AGN-starburst galaxy NGC~6764 with the VLBA at 1.6 and 4.9 GHz. The VLBA observations clearly detect a ``core-jet'' structure with a {\it possible} weak counterjet component
at 1.6 GHz. There is a tentative detection of the core and jet at 4.9 GHz, but it requires further confirmatory observations.
\item[2.] We derive upper limits to the 1.6$-$4.9 GHz radio spectral index for the core and jet, which are $>0.6$ and $>0.3$, respectively. Taken along with the high brightness temperature of 
$\sim10^7$ K for the core region, this suggests that the radio emission comes from a synchrotron jet.
\item[3.] The parsec-scale jet, at a P.A. of $\sim25^\circ$, appears to be 
pointing closely toward the region with the relatively flat radio and harder X-ray spectrum in the western edge of the southern bubble. A connection between the radio emission on parsec- and sub-kpc scales would indicate that the radio bubbles in NGC~6764 are indeed powered by an AGN jet. Furthermore, it would require the AGN jet to be curved.
\item[4.] We find that the curved (or misaligned) jet can be fitted with a precessing jet model from the 
 parsec to sub-kpc scales. However, not all of the radio emission in NGC~6764 can be fitted by a precessing or curved radio jet, as was successfully modeled for Mrk~6. The jet appears to be disrupted beyond sub-kpc scales, possibly due to the loss of momentum by interaction and/or entrainment of the ISM, leading eventually to the creation of lobe-like bubbles.
This is consistent with the copious X-ray emission observed co-spatial with the radio emission in NGC~6764 \citep{Croston08}.
\item[5.] Following the suggestion of a compact-radio-luminosity dichotomy in AGNs and starbursts   \citep{Kewley00}, we find that even though the extrapolated 2.3 GHz luminosity of NGC~6764 ($\sim5\times10^{19}$~W~Hz$^{-1}$) places it in the starburst category, the high brightness temperature and morphology strongly support the picture of an AGN in NGC~6764.  
\item[6.] The jet energetics suggest that it would take 12$-$21 Myr for the jet to inflate the (southern) bubble. This timescale corresponds roughly to the starburst episode that took place in NGC~6764 about 15$-$50 Myr ago. This could point to a close connection between the ejection of the AGN jet and the starburst episode, which might have resulted from a common accretion event in the galaxy.
\item[7.] NGC~6764 seems to present a convincing example of an AGN jet that is closely associated with the kpc-scale radio bubbles. It is tantalizing to invoke a correspondence between this picture and that proposed for powerful cluster radio galaxies like Perseus~A 
\citep[e.g.,][]{Dunn06,Falceta10}.
\end{enumerate}

\acknowledgments
We thank the referee for helpful suggestions which have improved this paper significantly.
We acknowledge the technical help provided by Joan Wrobel in scheduling the VLBA observations. This work is partially supported by NASA grant G08-9108X.
This research has made use of the NASA/IPAC Extragalactic Database (NED) which is operated by the Jet Propulsion Laboratory, California Institute of Technology, under contract with the National Aeronautics and Space Administration. M.J.H. thanks the Royal Society for support.

\bibliographystyle{apj}
\bibliography{ms}

\newpage

\begin{figure}
\centering{
\includegraphics[width=6.4cm]{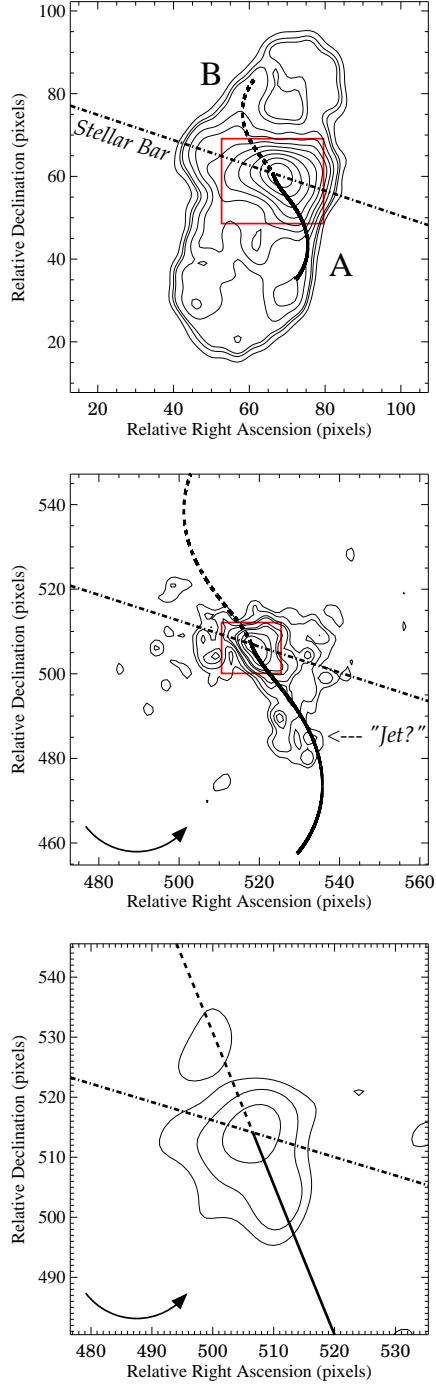}}
\caption{\small Precessing jet model can be fitted to the radio emission from parsec- to sub-kpc scales. (top) VLA A-array 1.4~GHz image (pixel size = 0.2$\arcsec$ = 31 pc);
(middle) VLA A-array 4.86~GHz image (pixel size = 0.1$\arcsec$ = 15.4 pc); and
(bottom) VLBA image at 1.6~GHz (pixel size = 0.4~$mas$ = 0.06 pc).
The peak radio emission is at R.A. = 19h 08m 16.432s, decl. = 50d 55m 59.525s.
The red box indicates the region zoomed in on the panel below, while the curved arrow indicates the sense of the galaxy rotation. Image contour levels are
(top) 0.25, 0.4, 0.5, 0.8, 1.13, 1.6, 2.25, 3.2, 4.5, 6.4, 9.0 mJy~beam$^{-1}$,
(middle) 0.08, 0.125, 0.2, 0.25, 0.4, 0.5, 0.8, 1.12, 1.6, 2.25, 3.2, 4.5, 6.4 mJy~beam$^{-1}$, and
(bottom) 0.18, 0.26, 0.37 mJy~beam$^{-1}$.
}
\end{figure}

\end{document}